\documentclass[twocolumn,secnumarabic,amssymb, nobibnotes, aps, prd, superscriptaddress]{revtex4-2}

\usepackage{pifont}
\usepackage[english]{babel}
\usepackage[utf8]{inputenc}
\usepackage[tbtags]{mathtools}
\usepackage{relsize}
\usepackage{environ}

\usepackage{helvet}									
\usepackage[T1]{fontenc}
\usepackage{lipsum}
\usepackage{graphicx}
\usepackage{dcolumn}
\usepackage[tight]{subfigure}
\usepackage{dsfont}
\usepackage{verbatim}
\usepackage{units}
\usepackage[usenames,dvipsnames]{color}
\usepackage{bm}
\usepackage{algorithmicx}
\usepackage{algpseudocode}
\usepackage{braket}
\usepackage{mathrsfs}
\usepackage{enumerate} 
\usepackage[normalem]{ulem}
\usepackage{parskip}

\definecolor{myBlue}{RGB}{8,81,156}
\definecolor{myRed}{RGB}{164,16,52}
\definecolor{myGreen}{RGB}{0,90,50}
\usepackage{hyperref}
\hypersetup{colorlinks=true,linktoc=all,linkcolor=myBlue,breaklinks=true,citecolor=myBlue,urlcolor=myBlue}
\addto\captionsenglish{\renewcommand{\figurename}{FIG.}}

\renewcommand{\vec}[1]{\mathbf{#1}}

\newcommand{\vl}{v_\text{L}}

\renewcommand{\vr}{v_\text{R}}

\renewcommand{\to}{\tilde{0}}
\newcommand{\op}[1]{\hat{ #1}}
\newcommand{\ops}[1]{\skew{4}\hat{ #1}}

\newcommand{\half}{\dfrac{1}{2}}

\newcommand{\Ham}{{\mathcal{H}}}

\newcommand{\acosh}{\,\text{arcosh}\,}

\newcommand{\ie}{i.e.,}

\newcommand{\cf}{\emph{cf.}}
\newcommand{\eg}{e.g.,}

\newcommand{\proj}[1]{ \ket{#1} \bra{#1}}

\newcommand{\mybar}[1]{\,\overline{\!{#1}}} 

\usepackage{microtype}

\begin{document}

\title{Electromotive force in driven topological quantum circuits}%

\author{Ahmed Kenawy}%
\affiliation{Peter Gr\"{u}nberg Institut, Forschungszentrum J\"{u}lich, D-52425 J\"{u}lich, Germany}
\author{Fabian Hassler}
\affiliation{Institute for Quantum Information, RWTH Aachen University, 52056 Aachen, Germany}
\author{Roman-Pascal Riwar}
\affiliation{Peter Gr\"{u}nberg Institut, Forschungszentrum J\"{u}lich, D-52425 J\"{u}lich, Germany}



\begin{abstract}

Time-dependent control of superconducting quantum circuits is a prerequisite for building scalable quantum hardware. The quantum description of these circuits is complicated due to the electromotive force (emf) induced by time-varying magnetic fields. Here, we examine how the emf modifies the fractional Josephson effect. We show that a time-varying flux introduces a new term that depends on the geometry of both the circuit and the applied magnetic field. This term can be probed via current and charge measurements in closed-loop and open-circuit geometries. Our results refine the current understanding of how to properly describe time-dependent control of topological quantum circuits.

\end{abstract}

\maketitle

\section{Introduction}

Superconducting circuits are one of the prime candidates for realizing large-scale quantum computers~\cite{Arute2019,Wu2021}. The behavior of these circuits can be captured by quantum circuit theory, a concise set of rules to construct a Hamiltonian by reducing circuits to lumped elements whose phase and charge are canonically conjugate variables~\cite{Devoret2017,Burkard2004,Ulrich2016}. Circuit quantization has been successfully applied to various devices such as quantum bits~\cite{Clarke2008,Schoelkopf2008,Devoret2013,Kjaergaard2019} and extended to include light-matter interactions, leading to circuit quantum electrodynamics (cQED)~\cite{Blais2004,Blais2007,Blais2020,Blais2021}.

Recently, it was shown that applying time-varying magnetic fields introduces several subtleties~\cite{You2019}, which asks for an extension of the standard lumped-element approach. The origin of these subtleties can be expressed as follows. For two circuit nodes~$a$ and~$b$ with a phase difference~$\phi\sim \smallint_a^b  \vec{A}(\vec{r}) \, \cdot \mathrm{d}\vec{r}$, the spatial distribution of the vector potential~$\vec{A} (\vec{r})$ is no longer an irrelevant gauge degree of freedom due to the induced electromotive force (emf)~$\sim \dot{\vec{A}}$. The correct gauge was identified in Ref.~\cite{Riwar2021} for general device and field geometries, and it was also applied to circuits with conventional Josephson junctions, where the Josephson energies are scalar quantities. An unanswered question, however, is how the emf changes Josephson effects whose energies exhibit an underlying matrix structure due to the presence of bound states.

One relevant example of the latter is the fractional Josephson effect, related to the presence of Majorana bound states. Majorana-based quantum circuits are a candidate for fault-tolerant quantum computation~\cite{Kitaev2003}. Topological superconductors hosting Majoranas may be realized, for example, using topological insulators~\cite{FuKane,Fu2009}, chains of magnetic adatoms~\cite{Nadj2013,Pientka2013}, proximitized semiconducting nanowires~\cite{Lutchyn2010,Oreg2010}, and hybrid semiconductor-superconductor planar heterostructures~\cite{Hell2017,Pientka2017}. Other potential platforms for realizing topological superconductors are discussed in~\cite{Alicea2012,Sato2017,Flensberg2021}. Irrespective of the physical implementation, a simple yet useful model for topological $p$-wave superconductors is the Kitaev chain, which consists of spinless electrons hopping between the sites of a finite one-dimensional lattice~\cite{Kitaev2001,Leijnse2012}. Given the prospect of fault-tolerant quantum computing, a better understanding of time-dependent control of Majorana circuits is required for implementing protected quantum gates~\cite{vanHeck2012,Hyart2013}.

In this paper, we study time-dependent driving of two weakly coupled Kitaev chains, placed on an interrupted superconducting loop that encloses a time-dependent magnetic flux. We derive a low-energy description of the resulting fractional Josephson effect, which arises because of the overlap between the two adjacent Majoranas of the coupled chains. We show that, in the presence of a time-dependent magnetic field, the fractional Josephson effect is modified by a new term that is a function of the time derivative of the phase difference between the two chains and that depends on both the geometry of the circuit and the profile of the external magnetic field. 

We further examine how this correction term scales with the parameters of the Kitaev chains and the spatial symmetry of the system, concluding that this term can be comparable to the Josephson energy for an asymmetric system. We explore the relevance of this new term in two contexts. In a loop geometry, it is equivalent to the linear response due to a low-frequency drive of the magnetic field. In an open-circuit geometry (\ie~cutting the loop to form two islands), it is equivalent to an additional charge induced on the chains, resulting in nonzero fluctuations of the number of Cooper pairs stored on a capacitor even in the regime of large charging energies. In this regime, while the Cooper-pair transport is energetically inhibited, charges in units of~$e$ (with~$e$ being the elementary charge) may still hop between the Majorana bound states, thereby yielding nonzero fluctuations. Overall, the measurable effects of this new term in both the loop and open-circuit geometries corroborate the relevance of the general gauge-fixing procedure, proposed in Ref.~\cite{Riwar2021}, to obtain the correct behavior of driven Majorana quantum circuits.


This paper is organized as follows. Section~\ref{model} introduces the model of two weakly coupled Kitaev chains and discusses the connection between time-dependent basis changes and the gauge degrees of freedom in the presence of a time-dependent magnetic field. In Sec.~\ref{low-energy-description}, we describe the low-energy physics of the fractional Josephson effect and show how the new correction term due to the electromotive force (emf) emerges. Section~\ref{current_measurement_sec} relates this term to the imaginary part of the linear response function of the current measured across the weak link. The dependence of this term on the parameters of the Kitaev chain is detailed in Sec.~\ref{irrotational_gauge_section} for several configurations of the external magnetic field. Next, Sec.~\ref{circuit} incorporates the effective Hamiltonian of the fractional Josephson effect into a minimal circuit and examines how the new term changes the behavior of the circuit.

\section{Setup}\label{model}

\begin{figure}[t]
	\centering
	\includegraphics[scale=1]{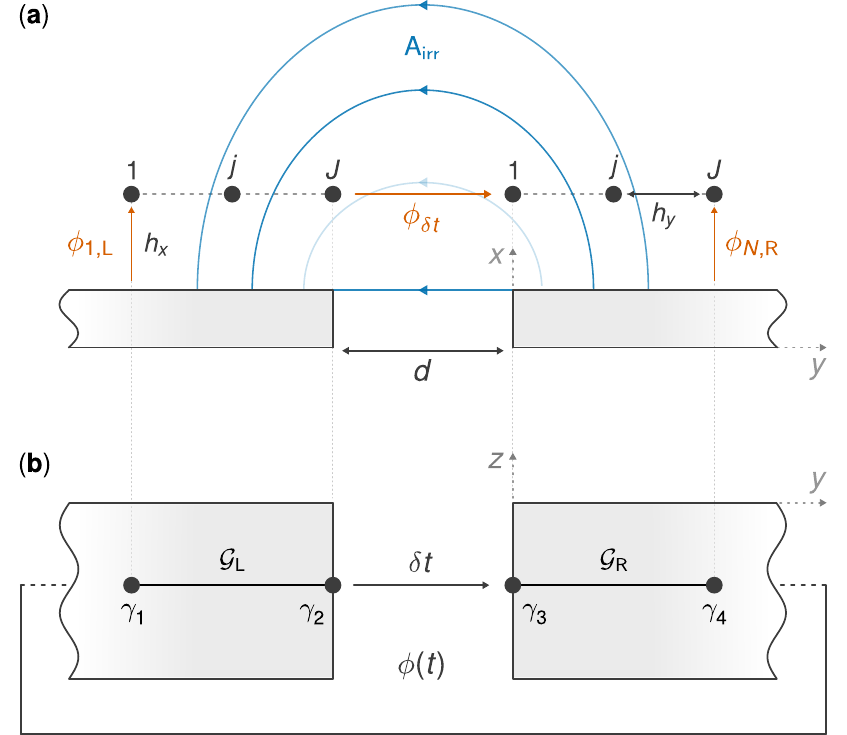}
	\caption{Two tunnel-coupled topological superconducting wires, modeled as two Kitaev chains of~$J$ sites each.~(\textbf{a})~Side view of the two chains on top of two coplanar superconducting plates, separated by a distance~$d$. The blue contours, perpendicular to the surface of the superconductor, represent the irrotational vector potential~\eqref{irrotational_vector_potential}. An electron acquires a phase~$\phi_{j, \alpha}$ as it hops to site~$j$ from the superconductor underneath with~$\alpha = \text{L},\text{R}$.~(\textbf{b})~Top view of the superconducting plates whose two far ends interconnect to form a loop threaded by an external magnetic field, resulting in a total phase difference~$\phi$ between the two superconductors. This phase difference~$\phi  = \phi_{ J,  \text{L}} + \phi_{\delta t} - \phi_{ 1,  \text{R}}$, with~$\phi_{\delta t}$ being the phase drop across the weak link. Here, the two chains are represented by the four Majoranas~$\{ \gamma_1, \gamma_2, \gamma_3, \gamma_4 \}.$ The profile of the phase drop in the chains is captured by the operators~$\mathcal{G}_\text{L}$ and~$\mathcal{G}_\text{R}.$} 
	\label{figure_model}
\end{figure}

We begin by reviewing the model for a flux-biased Majorana junction and discuss the relevant aspects of the gauge degrees of freedom. We consider two tunnel-coupled topological superconductors with a phase bias, as depicted in Fig.~\ref{figure_model}. A model for one-dimensional topological superconductors is provided by the Kitaev chain~\cite{Kitaev2001}. The two coupled chains are described by the Hamiltonian~$	H (\phi) = H_0  + H_\text{T} (\phi)$ with
\begin{equation} 
	H_0 = \sum_\alpha 	H_{\text{K},\alpha} ,
\end{equation}
where the subscript~$\alpha$ denotes either the left~($\alpha = \text{L}  $) or the right~($\alpha = \text{R}  $) chain. For a chain of~$J$ sites, the Kitaev Hamiltonian with nearest-neighbor hopping and pairing between adjacent sites is written as~\cite{Kitaev2001,Leijnse2012}
\begin{align}\label{Kitaev_Hamiltonian}
	H_{\text{K},\alpha} = &-\mu \sum_{j}^{J} c_{j, \alpha}^\dagger c_{j, \alpha}^{} - t  \sum_{j}^{J- 1} \Big(  c^\dagger_{j+1, \alpha} c_{j, \alpha}^{} + \text{H.c.} \Big)  \nonumber \\	&+ \Delta  \sum_{j}^{J - 1} \Big( c_{j+1, \alpha}^{} c_{j,\alpha}^{}
	+ \text{H.c.} \Big),
\end{align}
where~$\mu$ is the chemical potential,~$t$ is the hopping amplitude, and~$\Delta$ is the pairing potential. Throughout this work, we consider the Kitaev chains in the topological regime~$ |\mu| \le 2 t $. The fermionic operator~$c_{j, \alpha}$ annihilates an electron at site~$j$ of chain~$\alpha$. Here, the parameter~$\Delta$ is real, and the phase difference between the two Kitaev chains is included in the tunneling Hamiltonian
\begin{equation}\label{tunneling_link_gauge}
	H_\text{T}(\phi) = - \delta t \, \left(  e^{i \phi} \, c_{1, \text{R}}^\dagger c_{J,\text{L}}^{} +   e^{-i \phi} \, c_{J, \text{L}}^\dagger c_{1, \text{R}}^{} \right),
\end{equation}
which couples the last site of the left chain to the first site of the right chain with~$\delta t \ll \ t$. The phase difference~$\phi = \phi_\text{R} - \phi_\text{L} $ is acquired by an electron as it hops over the weak link due to the flux enclosed by the superconducting loop (Fig.~\ref{figure_model}). Alternatively, we can write Eq.~\eqref{tunneling_link_gauge} in terms of the phase acquired by a Cooper pair, henceforth denoted by~$\varphi$ (with~$\phi = \varphi /2$), as discussed in Sec.~\ref{circuit}. Contrary to the ordinary Josephson effect, the tunneling Hamiltonian allows for hopping of individual electrons\textemdash not Cooper pairs\textemdash across the junction, leading to the fractional Josephson effect~\cite{FuKane}.

In this work, we study time-dependent control of the magnetic flux. Therefore, we have to worry about the spatial distribution of the phase\textemdash that is, the allocation of the phase difference~$\phi$ to the weak link and the two Kitaev chains\textemdash similar to Refs.~\cite{You2019,Riwar2021}, which can be explained as follows. For a time-independent flux~($\dot{\phi}=0$), the choice of where the phase difference~$\phi$ enters the Hamiltonian is immaterial. Instead of having the phase difference attached locally to the weak link, as in Hamiltonian~\eqref{tunneling_link_gauge}, one may choose any other phase distribution (local or nonlocal), as long as an electron acquires a total phase~$\phi$ when going from the left to the right superconductor. This entire family of Hamiltonians can be related to~$H$ via the unitary transformation
\begin{align} \label{unitary}
	\mybar{H} &= U H U^\dagger \nonumber  \\
	&= e^{i\phi G} \, H \, e^{-i\phi G},
\end{align}
with
\begin{equation}\label{first_def_G}
	G \equiv \sum_{j,\alpha} \, \dfrac{\phi_{j,\alpha}}{\phi} \: c_{j,\alpha}^\dagger c_{j, \alpha},
\end{equation}
where~$\phi_{j,\alpha}$ denotes the phase acquired at each site of the Kitaev chain in the basis of the Hamiltonian~$\mybar{H}$. Here, we normalize the site phases by the total phase difference~$\phi$ such that the operator~$G$ is independent of~$\phi$. The unitary transformation~$U$ redistributes the phase across the sites of the left and right chains (\ie~across the fermionic operators~$c_{j,\alpha}$). The different bases defined by~$U$ correspond to gauge choices because they do not change the eigenspectrum of the Hamiltonian. But, for a time-dependent flux~$(\dot{\phi}\neq 0)$, the unitary transformation itself depends on time and yields the additional term~$ - i \hbar U^\dagger  \dot{U}$ in the Schrödinger equation. For this reason, a time-dependent change of basis via a unitary transformation is not equivalent to a gauge transformation. Therefore, when constructing a Hamiltonian for a driven quantum circuit, one must ensure that the phase distribution entering the Hamiltonian corresponds to an actual gauge choice.


Reference~\cite{Riwar2021} proposed a recipe to obtain the correct time-dependent Hamiltonian by calculating the vector potential in the so-called irrotational gauge, which combines the Coulomb gauge~$	\bm{\nabla} \cdot \vec{A}_\text{irr} = 0$ with the boundary condition
  \begin{equation}\label{key}
\vec{n}_{\perp} \times \vec{A}_\text{irr} \big|_{\mathcal{S}} = 0,
 \end{equation}
 with the unit vector~$\vec{n}_{\perp} $ normal to the surface~$\mathcal{S}$ of the superconductor, for geometries where the London penetration depth is an irrelevant length scale. This gauge ensures that there are no terms in the Hamiltonian proportional to~$\dot{U}$ (or, for the Kitaev model discussed here, to~$\dot{\phi}$) by including the entirety of the emf induced by the time-dependent magnetic field\textemdash that is,~$\bm{\nabla} \times	\vec{E}_{\dot{{B}}}=\dot{\vec{B}}$\textemdash in the irrotational vector potential
\begin{equation}\label{induced_E}
	\vec{E}_{\dot{{B}}} = -  \dot{\vec{A}}_\text{irr}.
\end{equation}

After deriving the correct time-dependent Hamiltonian~$\mybar{H}$, one can apply a unitary transformation to revert to the basis of the Hamiltonian~$H$ (hereafter the link basis as it allocates the entire phase difference to the weak link). In the link basis, the dynamics of the two coupled chains is governed by the Hamiltonian
\begin{equation}\label{key}
	H - i\hbar U^\dagger \dot{U} = H + \hbar \dot{\phi} G,
\end{equation}
seeing that we can write the operator~$G$ in Eq.~\eqref{first_def_G} as
\begin{equation}\label{G_definition}
	G = -i U^\dagger \partial_\phi U.
\end{equation}
As previously mentioned, the operator~$G$ accounts for the distribution of the phase difference across the sites of the Kitaev chains and uniquely defines the Hamiltonian~$\mybar{H}$.  Later in Sec.~\ref{irrotational_gauge_section}, we evaluate the site phases~$\phi_{j, \alpha}$ and the corresponding~$G$ for an explicit circuit model with simple yet realistic assumptions on circuit geometry and the magnetic field. As argued in Ref.~\cite{Riwar2021}, the spatial distribution of the vector potential and, in turn, the site phases hinge on the profile of the magnetic field.

To conclude this section, we assess the difference between the two Hamiltonians~$H$ and~$\mybar{H}$ in terms of current measurements. In the link basis, the current across the weak link is defined as
\begin{equation}\label{current_link}
	I_0 =  - \dfrac{e}{\hbar} \partial_\phi H (\phi) = - \dfrac{e}{\hbar} \partial_\phi H_\text{T} (\phi),
\end{equation}
where~$e > 0$ is the elementary charge. Because the Kitaev Hamiltonian~\eqref{Kitaev_Hamiltonian} does not depend on~$\phi$, the current~$I_0$ only depends on the tunneling Hamiltonian. In the basis of the Hamiltonian~$\mybar{H}$ (hereafter the irrotational basis), there exists a current that can be defined as
\begin{equation}\label{current_irrotational}
		\mybar{I} = - \dfrac{e}{\hbar} \, \partial_\phi \mybar{H}(\phi)  = \mybar{I}_0 +  e \dot{\mybar{G}} \\
\end{equation}
where~$\dot{\mybar{G}} = (i / \hbar) [\mybar{H} , \mybar{G}]$ and~$\mybar{I}_0$ is the current across the weak link in the irrotational basis~(see Appendix~\ref{current_irrotational_appendix} for the derivation). The operator~\eqref{current_irrotational} can be interpreted as the current in response to a weak, time-dependent magnetic field. This interpretation is motivated by replacing the phase~$\phi(t)$ by~ $\phi + \delta \phi(t)$ and approximating the Hamiltonian, up to first order in the fluctuation~$\delta \phi(t)$, by~$ \mybar{H}(\phi) + \delta \phi(t) \, \partial_\phi \mybar{H}(\phi)$. The time-dependent part of the current~\eqref{current_irrotational} depends on the phase distribution across the sites of the chain, captured here by the operator~$\mybar{G}$. Thus, we can interpret~$e \dot{\mybar{G}}$ as a displacement current and~$e \mybar{G}$ as the corresponding charge, which is zero for a magnetic field that couples exclusively to the weak link. The relation~\eqref{current_irrotational} will be relevant when computing the linear response function of the current~$\mybar{I}_0$ in Sec.~\ref{current_measurement_sec}.

\section{Effective low-energy model}\label{low-energy-description}
Before considering specific circuit geometries, it is helpful to reduce the complexity of the problem by deriving a low-energy approximation of the Hamiltonian. To that end, we start from the generally valid Hamiltonian~$H(\phi) + \hbar \dot{\phi} G$ and apply a Schrieffer-Wolff transformation, projecting to the low-energy subspace comprising the four degenerate ground states of the two Kitaev chains~\cite{Wolff1966}.

The Hamiltonian~$H$ can be cast into the form
\begin{equation} 
	\begin{split}
		H &=  \half \psi^\dagger \mathcal{H} \psi \\ &= \half    \begin{pmatrix}
			\psi^\dagger_\text{L} & 	\psi^\dagger_\text{R}  
		\end{pmatrix} 
		\begin{pmatrix}
		\Ham_{\text{K}, \text{L}}  & \mathcal{W}\\   \mathcal{W}^\dagger &  \Ham_{\text{K}, \text{R}}
	\end{pmatrix}
		\begin{pmatrix}
		\psi^{}_\text{L} \\	\psi^{}_\text{R}
	\end{pmatrix}   ,
	\end{split}
\end{equation}
with the column vectors~$\psi_\alpha = ( \psi_{1, \alpha}  , \hdots , \psi_{j, \alpha}, \hdots , \psi_{J, \alpha})^\text{T} $ where each element denotes a spinor~$\psi_{j, \alpha} = (c_{j, \alpha}^{},  c_{j, \alpha}^\dagger)^\text{T}$. The Hamiltonian of an isolated chain can be decomposed in terms of its eigenstates as
\begin{equation} 
	\Ham_{\text{K},\alpha} = \sum_{v_\alpha \ge0} \epsilon_{v_\alpha} \ket{v_\alpha}\bra{v_\alpha}  -\epsilon_{v_\alpha} \ket{\tilde{v}_\alpha}\bra{\tilde{v}_\alpha},
\end{equation}
where~$	\ket{v_\alpha} = \tau_x \ket{\tilde{v}_\alpha}$ and the Pauli matrix~$\tau_x$ acts on the particle and hole blocks. Each chain has two zero-energy ground states of opposite parity, an even state~$\ket{0}$ and an odd state~$\ket{\tilde{0}}.$ The tunneling matrix~$\mathcal{W}$ can be written as
\begin{equation}\label{w_structure}
	\mathcal{W}_{j,j^\prime} =  \delta_{j,J} \:  \delta_{j^\prime,{1}}  \: \begin{pmatrix}
		-\delta t  \, e^{-i\phi} & 0 \\0 & 		\delta t \, e^{i\phi}
	\end{pmatrix},
\end{equation}
so that it couples the rightmost site of the left chain to the leftmost site of the right chain. Finally, the operator~${G}$ takes the form
\begin{equation} 
	{G} =  \half \psi^\dagger \mathcal{G} \psi =\half   \begin{pmatrix}
		\psi^\dagger_\text{L} & 	\psi^\dagger_\text{R} 
	\end{pmatrix}
	\begin{pmatrix}
		\mathcal{G}_\text{L} & 0 \\  0& \mathcal{G}_\text{R}
	\end{pmatrix}
	\begin{pmatrix}
		\psi^{}_\text{L} \\	\psi^{}_\text{R}
	\end{pmatrix}.
\end{equation}
The matrices~$\mathcal{G}_\text{L}$ and~$\mathcal{G}_\text{R}$ depend on the profile of the external magnetic field, as detailed in Sec.~\ref{irrotational_gauge_section}.

By projecting onto the low-energy subspace described by the operator
\begin{equation} \label{projector_p}
		\mathcal{P}  = \sum_{v = 0, \tilde{0}} \bigg[ \begin{pmatrix}
			\ket{v_\text{L}} \bra{v_\text{L}} & 0 \\
				0 & 0
			\end{pmatrix} 
		+  \begin{pmatrix}
		0 & 0 \\
			0 & 	\ket{v_\text{R}} \bra{v_\text{R}} 
		\end{pmatrix} \bigg],
	\end{equation}
one can obtain the effective Hamiltonian
\begin{equation}\label{final_result_sw}
		\mathcal{H}_\text{eff} =  \mathcal{P} \mathcal{H}_\text{T} \mathcal{P}  
		-  \hbar \dot{\phi} \,  \mathcal{P}   \bigg(    \mathcal{H}_\text{T}   \dfrac{\mathcal{Q}}{\mathcal{H}_0}  \mathcal{G}   +  \mathcal{G}   \dfrac{\mathcal{Q} }{\mathcal{H}_0}  \mathcal{H}_\text{T}   \bigg)  \mathcal{P} , 
\end{equation}
which decouples the low-energy subspace from the high-energy subspace described by the projector~$\mathcal{Q} = 1 - \mathcal{P}$. The above result corresponds to a perturbative expansion up to first order in~$\delta t$ and~$ \dot{\phi}$ (Appendix~\ref{sw_appendix}).

Reverting to second quantization, the resulting low-energy Hamiltonian is 
\begin{equation}\label{effective_ham}
	H_\text{eff}  =  H_\text{M}(\phi) + \hbar \dot{\phi} \, {H}_\lambda (\phi).
\end{equation}
The first term corresponds to the well-known Hamiltonian that describes the fractional Josephson effect 
\begin{equation} 
	H_\text{M} (\phi) = i E_\text{M}  \cos \phi \: \gamma_\text{3}^{}  \gamma_\text{2}^{},
\end{equation}
where the two Hermitian operators~$\gamma_2$ and~$\gamma_3$ denote the adjacent Majoranas of the left and right chains, respectively (Fig.~\ref{figure_model}). Importantly, the time-dependent flux leads to a new term, which is our first central result,
\begin{equation} 
	{H}_\lambda (\phi) =  i \lambda \cos \phi \: \gamma_\text{3}^{}  \gamma_\text{2}^{},
\end{equation}
with the dimensionless coupling coefficient
\begin{equation}\label{lambda_def} 
\lambda   =    \Re \bigg\{  \dfrac{ \bra{0_\text{L}} \mathcal{G}_\text{L}^{}  \,  \Sigma_\text{L}    \,  \mathcal{W}  \ket{0_\text{R}} +   \bra{{0}_\text{L}}   \mathcal{W}  \, \Sigma_\text{R} \, \mathcal{G}_\text{R}^{}      \ket{0_\text{R}}}{\cos \phi } \bigg\} ,
\end{equation}
and
\begin{equation} 
	\Sigma_\alpha \equiv \sum_{v_\alpha > 0}  \dfrac{\ket{v_\alpha} \bra{v_\alpha}  - \ket{\tilde{v}_\alpha} \bra{\tilde{v}_\alpha}  }{\epsilon_{v_\alpha}} . 
\end{equation}
The coefficient~$\lambda$ does not depend on the phase~$\phi$ because the numerator is also proportional to~$\cos \phi$. It, however, depends on the interaction of the time-dependent magnetic field with the left and right chains, as expressed by~$\mathcal{G}_\text{L}$ and~$\mathcal{G}_\text{R}$. In Sec.~\ref{irrotational_gauge_section}, we compare several profiles of the magnetic field and explore how~$\lambda$ depends on the parameters of the chain\textemdash namely, the chemical potential~$\mu$, the hopping amplitude~$t$, and the pairing potential~$\Delta$. 

Importantly, because the tunneling Hamiltonian couples the last site of the left chain to the first site of the right chain, only Majoranas~$\gamma_2$ and~$\gamma_3$ interact, despite the nonlocal phase distribution in the irrotational gauge. Specifically, interaction terms that include Majoranas~$\gamma_1$ and~$\gamma_4$ are exponentially suppressed because, throughout this work, we consider Kitaev chains (\ie~nanowires) that are long enough so that the Majoranas at the two ends of each chain do not overlap. For shorter nanowires with overlapping Majoranas, there would be additional terms that couple the four Majoranas~$\gamma_{\{1-4\}}$ (\eg~the pairs~$\gamma_4 \gamma_1$,~$\gamma_4 \gamma_2$, and~$\gamma_3 \gamma_1$ interact with a coefficient proportional to~$ \delta_t \dot{\phi}$, and the pairs~$\gamma_2 \gamma_1$ and~$\gamma_4 \gamma_3$ with a coefficient proportional to~$\dot{\phi}$, as detailed in Appendix~\ref{sw_appendix}). Previous works have shown that such terms lead to new phenomena such as the Landau-Zener effect and dc Shapiro steps~\cite{Feng2018,Choi2020}. Strikingly, our work shows that, even in the ideal case where finite-size effects can be neglected and the system is topologically protected, a time-dependent drive can lead to new physics, evidenced by the correction to the fractional Josephson effect [Eq.~\eqref{effective_ham}].

\section{Current Measurements}\label{current_measurement_sec}

We have shown that, in the presence of a time-dependent magnetic field, the fractional Josephson effect is modified by a new term that depends on the time-derivative of the phase difference between two weakly coupled Kitaev chains and on the coefficient~$\lambda$. This section examines how the coefficient~$\lambda$ can be measured. Suppose there is a detector measuring the current~$I_0$ at the weak link~[Eq.~\eqref{current_link}]. Let us now consider the linear response of this current to a time-dependent drive of the magnetic flux in the form~$  \phi + \delta \phi (t) $. In the frequency domain, the Kubo formula reads~\cite{Kubo1966}
\begin{equation} 
	\braket{\mybar{I}_0(\omega)} =   	\braket{\mybar{I}_0(\omega)}_0 + \delta \phi(\omega)\,  \chi (\omega),
\end{equation}
where the susceptibility is defined as
\begin{equation} 
	\chi^{}(\omega) =  \dfrac{i}{e} \int_{-\infty}^{\infty} \mathrm{d} t \: e^{i\omega t } \, \Theta(t) \, \Braket{[ \mybar{I}_0(0), \mybar{I}(-t)  ]  }_0,
\end{equation}
which naturally depends on the current associated with the magnetic-field drive, as introduced in Eq.~\eqref{current_irrotational}.
This formula holds for a generic initial state, expressed by the expected value~$\braket{\bullet}_0$. Here, we consider a low-energy regime, where the relevant physics occur within a subspace that consists of the four degenerate ground states of the left and right chains. This low-energy subspace is projected onto by the operator~$P$, which takes the form~\eqref{projector_p} in the two-by-two space of the left and right chains. Now, to account for arbitrary initial states, we choose to define a susceptibility operator
\begin{equation} \label{sus_op}
	\op{\overline{\chi}}(\omega) =  \dfrac{i}{e} \int_{-\infty}^{\infty} \mathrm{d} t \: e^{i\omega t } \, \Theta(t) \, P [ \mybar{I}_0(0), \mybar{I}(-t)  ] P,
\end{equation}
which acts on said subspace. As detailed in Appendix~\ref{sus_appendix}, substituting with the form~\eqref{current_irrotational} of the current~$\mybar{I}$ and keeping only terms that are first order in the tunneling amplitude~$\delta t$ gives the operator
\begin{equation} 
	\op{\overline{\chi}}(\omega)  = - \mybar{P} \bigg(   \mybar{I}_0 \mybar{Q} \dfrac{i \mybar{H}_0}{{ \hbar \omega } - \mybar{H}_0} \mybar{G} + \mybar{G} \dfrac{i \mybar{H}_0}{ { \hbar \omega }- \mybar{H}_0}  \mybar{Q}  \mybar{I}_0  \bigg) \mybar{P}.
\end{equation}
The operator~$\op{\overline{\chi}} $ is non-Hermitian and hence its eigenvalues have real and imaginary parts. We can then decompose it into two operators, which lead to the real and imaginary parts of the susceptibility when taking the expected value,
\begin{equation}\label{hermitian_part}
	\op{\overline{\chi}}_r(\omega) = \dfrac{1}{2}  \Big[ \op{\overline{\chi}}  + \big(\op{\overline{\chi}}\big)^\dagger \Big],
\end{equation}
and
\begin{equation}\label{antihermitian_part}
	\op{\overline{\chi}}_i(\omega) =  \dfrac{1}{2i}  \Big[ \op{\overline{\chi}}  - \big(\op{\overline{\chi}}\big)^\dagger \Big].
\end{equation}
In the low-frequency limit and in the link basis, the first part can be written as
\begin{equation} \label{general_real_sus}
	  	\op{{\chi}}_r^{} (\omega)   = i {P} \Big[   {I}_0(\phi) {Q}  {G} - {G}  {Q}  {I}_0(\phi)  \Big] {P}.
\end{equation}
Following Appendix~\ref{sus_appendix},~Eq.~\eqref{general_real_sus} simplifies to its final form
\begin{equation} \label{gauge_invar_sus}
\op{{\chi}}_r^{} (\omega) = \dfrac{e}{\hbar} \bigg( 1 - \dfrac{\phi_{\delta t}}{\phi} \bigg)  \, i E_\text{M}  \cos \phi \, \gamma_{3} \gamma_{2},
\end{equation}
where~$\phi$ is the total phase difference between the two superconductors, and~$\phi_{\delta t}$ is the phase drop across the weak link in the basis of the Hamiltonian~$\mybar{H}$, obtained by integrating the irrotational vector potential along the junction. In the irrotational basis, the phase drop~$\phi_{\delta t}$ is a well-defined quantity, much like the vector potential in the London equations~\cite{tinkham}. Rotating to a different gauge changes~$\phi_{\delta t}$ and also results in an additional term in the Hamiltonian, so that the left-hand side of Eq.~\eqref{gauge_invar_sus} is unchanged (\ie~gauge invariant). The low-frequency part of the real susceptibility shows that the phase drop across the chains leads to a new current contribution, different from the current flowing through the junction. If the entire phase drop takes place across the weak link~($\phi_{\delta t} = \phi$), then~$ \op{{\chi}}_r^{}  = 0$ and this additional current vanishes.

As for the imaginary part of the susceptibility, Eq.~\eqref{antihermitian_part} in the link basis reduces to
\begin{equation}\label{imag_sus}
\op{{\chi}}_i^{} (\omega) 	= { \hbar }\omega P \bigg(  I_0(\phi)  \dfrac{Q}{H_0} G + G \dfrac{Q}{H_0}  I_0(\phi) \bigg) P.
 \end{equation}
Since only the measured current~$I_0$ is a function of~$\phi$, Eq.~\eqref{imag_sus} can be cast into the differential form
\begin{equation} 
	I_\lambda (\phi) =  	\op{{\chi}}_i^{} (\omega)   	 =  e  \omega   \, \dfrac{\partial  {H}_\lambda}{\partial \phi},
\end{equation}
which, in a broad sense, can be regarded as a variant of the fluctuation-dissipation theorem stating that the time-dependent correction term in the effective Hamiltonian~\eqref{effective_ham} results in a new current contribution that is proportional to the low-frequency part of the imaginary susceptibility. In Sec.~\ref{circuit}, we propose an alternative way to probe the coefficient~$\lambda$ by applying Hamiltonian~\eqref{effective_ham} to an open-circuit geometry.

\section{Irrotational Gauge for two weakly coupled Kitaev chains}\label{irrotational_gauge_section}
We now embark on computing the irrotational vector potential for the model of two coupled Kitaev chains placed on an interrupted superconducting loop to find explicit expressions for the operator~$G$ and the resulting coefficient~$\lambda$. To simplify the problem, we assume the following. First, since the circumference of the loop is much larger than the length of the chains, the problem can be modeled as two coplanar superconducting plates separated by a distance~$d$ and whose width along the~$z$ direction is much larger than~$d$ (Fig.~\ref{figure_model}). These two plates are connected at the far end to form a loop that is threaded by the external magnetic field. Second, we assume that the field is nonzero only within the loop\textemdash that is, it touches neither the chains nor the superconducting plates. Third, we assume that only the superconducting contacts\textemdash not the Kitaev chains\textemdash screen the induced electric field~$\vec{E}_{\dot{B}}$ in Eq.~\eqref{induced_E} and, accordingly, we interpret~$e G$ as an induced charge on the chains. This assumption allows us to neglect the presence of the chains when computing the irrotational vector potential. Later, in Sec.~\ref{circuit}, we revisit this assumption to discuss the relevance of the~$\lambda$ term in an open-circuit geometry.

\begin{figure}[t]
	\centering
	\includegraphics[scale=1]{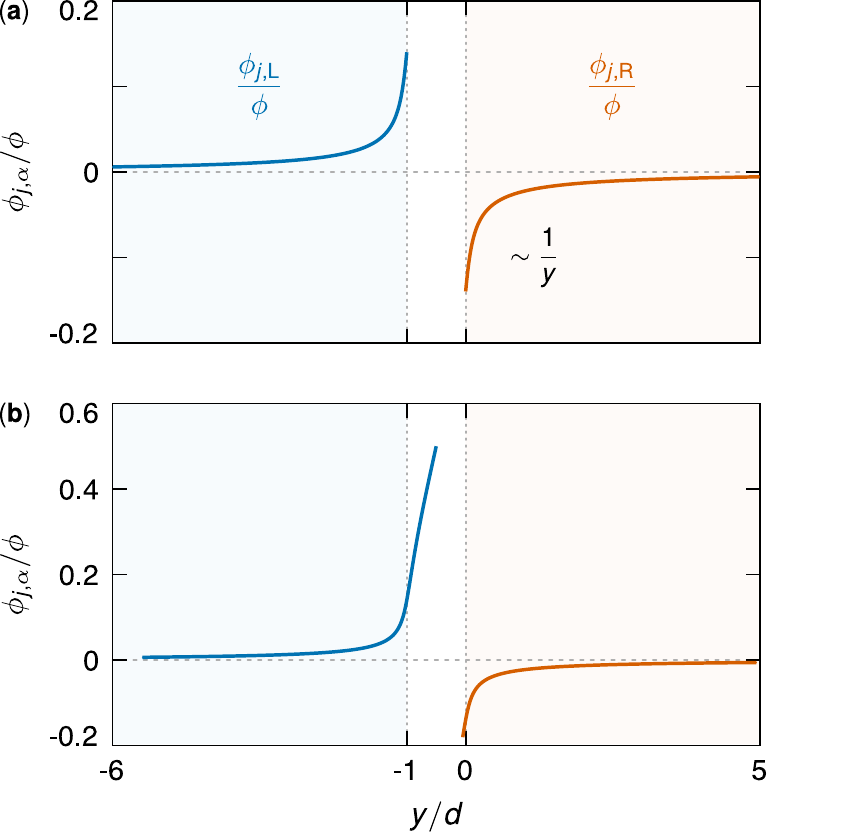}
	\caption{Phases acquired by electrons at the sites of the two Kitaev chains, based on Eq.~\eqref{general_phase_symmetric}. The two chains are on top of two coplanar superconducting plates (blue and red shaded areas), separated by a distance~$d$ (Fig.~\ref{figure_model}). The site phases~$\phi_{j, \alpha} $ are normalized by the phase difference~$\phi$ between the two superconductors.~(\textbf{a})~A symmetric configuration where the last site of the left chain and the first site of the right chain are aligned with the edges of the superconductors. The phase profile decays as~$1 /y $.~(\textbf{b})~An asymmetric configuration where the edges of the chains and the superconductors do not coincide. Here, five sites of the right chain and~$50$ sites of the left chain fall within the gap between the two superconductors. The phase drop~$\phi_{\delta t}$ across the weak link is larger in the symmetric configuration~$(\phi  = \phi_{ J,  \text{L}} + \phi_{\delta t} - \phi_{ 1,  \text{R}})$. Parameters used:~$J = 500$,~$d = 1$,~$h_x = 0.1 $, and~$h_y = 0.01 $.}
	\label{figure_phase}
\end{figure}

In the irrotational gauge, the vector potential~$\vec{A}$ must satisfy three conditions~\cite{Riwar2021}. First, its curl must be zero. Second, at the surface of the superconductor, its parallel component must be zero, as depicted by the blue contours in Fig.~\ref{figure_model}(\textbf{a}). Third, its normal component must integrate to zero over the surface of the superconductor, implying zero net surface charge. In fact, in conjunction with the magnetic field being zero in the bulk of the superconductors, the problem of finding the irrotational vector potential for a fixed phase bias can be mapped to the standard electrostatic problem of a coplanar capacitor with a voltage bias, which has been solved, for example, in Ref.~\cite{Riwar2018} using conformal mapping. Analogous to the electrostatic problem, the irrotational vector potential is
\begin{equation} \label{irrotational_vector_potential}
	\vec{A}_\text{irr}(x, y) =  \dfrac{\hbar \phi}{e\pi} \: \bm{\nabla} \, \Im \Bigg\{ \acosh \bigg( \dfrac{2y + 2i x }{d} + 1 \bigg) \Bigg\},
\end{equation}
where~$\phi$ is the overall phase difference between the left and right superconductors~($ \phi_\text{R} - \phi_\text{L} = \phi $). The vector potential can be integrated to compute the phase difference between the discrete sites of the Kitaev chain. In particular, an electron acquires the Peierls phase
\begin{align}\label{general_phase_symmetric}
		\phi_{j, \alpha}^{} & =  -   \dfrac{e}{\hbar} \int_{(0, y_{j, \alpha})}^{(h_x, y_{j, \alpha})}   \vec{A}_\text{irr} (x,y) \cdot \mathrm{d}\vec{l}  \nonumber  \\
		&= -\dfrac{\phi}{\pi} \, \Big[ f \big(h_x,  y_{j, \alpha}^{}  \big)- f\big(0, y_{j, \alpha}^{}    \big)   \Big]
\end{align}
as it hops to site~$j$ of chain~$\alpha$ from the superconductor underneath, with~$h_x$ as the height of the chain above the superconductor. The positions of the chain sites are
\begin{equation}\label{y_L}
	y_{j, \text{L}} \equiv -d - h_y (J - j),
\end{equation}
and
\begin{equation}\label{y_R}
	y_{j, \text{R}} \equiv h_y (j - 1),
\end{equation}
with~$h_y$ as the intersite separation. The function~$f(x,y)$ in Eq.~\eqref{general_phase_symmetric} is defined as
\begin{equation}
		f \equiv \arg \left( 2 i x + 2y + d + 2 \sqrt{ y + i x } \sqrt{y + d + ix}    \right).
\end{equation}
When evaluating the phases in Eq.~\eqref{general_phase_symmetric}, the~$y$ coordinate of the lower limit of the integral is irrelevant because the entire superconductor beneath each chain is at the same phase, as evidenced by the vanishing parallel component of the irrotational vector potential at the surface of the superconductor. The phase drop along the left and right chains decays as~$1/y$ starting from the gap between the two superconductors, as depicted in Fig.~\ref{figure_phase}(\textbf{a}). The phase drop across the weak link itself is
\begin{equation} \label{link_phase}
	\phi_{\delta t}^{} =  -\dfrac{\phi}{\pi} \, \Big[ f \big(h_x,  0\big)- f\big(h_x,  -d   \big)   \Big] .
\end{equation}

Using the phases in Eq.~\eqref{general_phase_symmetric}, the Hamiltonian in the irrotational basis can be written as
\begin{equation}
	\mybar{H}(\phi)  = \sum_\alpha \mybar{H}_{\text{K}, \alpha} (\phi) + \mybar{H}_\text{T}(\phi),
\end{equation}
with the Kitaev Hamiltonian
\begin{align}\label{Ham_irr}
		\mybar{H}_{\text{K},\alpha} = &-\mu \sum_{j}^{J} c_{j, \alpha}^\dagger c_{j, \alpha}^{}   \nonumber \\ & + \sum_{j}^{{J}-1} \Big[ - t  \Big(e^{i ( \phi_{ j+1, \alpha} -\phi_{ j, \alpha} ) } \, c_{j+1, \alpha}^\dagger c_{j, \alpha}^{} +  \text{H.c.} \Big) \nonumber \\  &  + \Delta  \Big( e^{-i ( \phi_{ j+1, \alpha} +\phi_{ j, \alpha} ) } \, c_{j+1, \alpha}^{}  c_{j,\alpha}^{} + \text{H.c.}   \Big) \Big] ,
\end{align}
and the tunneling Hamiltonian
\begin{equation}
	\mybar{H}_\text{T}= - \delta t   \left( e^{i\phi_{\delta t}^{}}  \, c_{{1}, \text{R}}^\dagger c_{J, \text{L}}^{} +  e^{-i \phi_{\delta t} }  \, c_{J, \text{L}}^\dagger c_{{1},\text{R}}^{} \right).
\end{equation}
In contrast to the tunneling Hamiltonian~\eqref{tunneling_link_gauge}, the phase difference between the two superconductors is not entirely allocated to the weak link but distributed along the chains. The actual phase drop across the weak link is less than the total phase difference~$\phi$, and it depends on the geometry of both the chains and the superconductors. Specifically, it increases as the height~$h_x$ of the chains above the superconductors decreases, taking the value~$\phi$ at~$h_x = 0$. And it increases as the separation~$d$ between the two superconductors increases, asymptotically approaching~$\phi$ as~$d$ tends to infinity.

In line with Sec.~\ref{low-energy-description}, we can now evaluate the Hermitian operator~$G$, which has left and right contributions
\begin{equation}
		G = G_\text{L} + G_\text{R}  = \sum_{j, \alpha}   \dfrac{\phi_{j,\alpha}}{\phi}    \: c_{j, \alpha}^\dagger c_{j, \alpha}^{} ,
\end{equation}
with the site phases~$\phi_{j,\alpha}$ incorporating the geometry of the chains and the profile of the magnetic field. In the single-particle picture, the operator~$\mathcal{G}_\alpha$ takes the form 
\begin{equation} 
	\mathcal{G}_\alpha^{j, j^\prime} = \delta_{j,j^\prime}  \,  \dfrac{\phi_{j,\alpha}}{\phi}   \,  \tau_z ,
\end{equation}
where~$\tau_z$ is the Pauli matrix acting on the particle and hole subspaces.

In the symmetric setup in Fig.~\ref{figure_model}(\textbf{a}), the phases~$\phi_{j, \alpha}$ are an odd function around~$y = -d/2$, and the phase profiles of the left and right chains are related by
\begin{equation}\label{filed_symmetry}
	\mathcal{G}_\text{L}^{j, j} = - \mathcal{G}_\text{R}^{J - j +1, J- j + 1} .
\end{equation}
Substituting with the above relation into Eq.~\eqref{lambda_def} gives~$\lambda $ equals zero. That is to say, the time-dependent correction~$ \hbar\dot{\phi} {H}_\lambda (\phi)$ to the fractional Josephson effect is zero for a symmetric setup. This behavior follows from the general form of the correction term, namely~$\dot{\phi} \cos \phi $. Because this term is an odd function of the total phase difference~$\phi$, it must be zero if the left and right chains are identical and thus interchangeable. Conversely, the Josephson term~$H_\text{M}(\phi)$ is an even function of~$\phi$ and thus nonzero for a symmetric setup.

For a nonzero~$\lambda$, one must break the symmetry between the left and right chains, either by choosing different parameters~($\mu$,~$t$, or~$\Delta$) for each chain, or by applying an external magnetic field that leads to an asymmetric distribution of the phase drop.

\begin{figure}[t]
	\centering
	\includegraphics[scale=1]{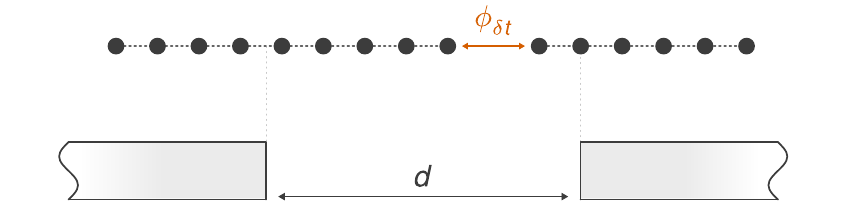}
	\caption{Two coupled Kitaev chains, placed on top of two superconducting plates. Unlike the symmetric setup in Fig.~\ref{figure_model}, the placement of the weak link ensures an asymmetric phase distribution over the left and right chains (\cf~Fig.~\ref{figure_phase}). This asymmetry yields a nonzero~$\lambda$ in the effective Hamiltonian~\eqref{effective_ham}.}
	\label{asymetirc_chain}
\end{figure}

Let us here, however, discuss another relevant source of spatial asymmetry: an asymmetric placement of the weak link. To that end, instead of having all sites of the chains in direct electric contact with the superconducting plates~(Fig.~\ref{figure_model}), we use an experimentally more realistic setup with the two superconductors connected by a bridge consisting of many sites~(Fig.~\ref{asymetirc_chain}). One can then choose to place the weak link between two intermediate sites to be closer to one of the two plates, thereby creating an asymmetric phase profile. An example of an asymmetric phase profile is shown in Fig.~\ref{figure_phase}(\textbf{b}), where, because the weak link is now closer to the right superconductor, the majority of the phase difference~$\phi$ is assigned to the left chain~($\phi_{J, \text{L}} > \phi_{1, \text{R}}$). And the phase drop across the weak link is smaller than the symmetric configuration.

While it is possible to modify the Hamiltonian~\eqref{Ham_irr} to include the extra sites between the two superconducting plates and the asymmetric placement of the junction, we simplify the problem and incorporate the asymmetry in the existing the Hamiltonian by modifying only the phase profile. For an increasingly asymmetric junction, all the phase drop will be allocated to the left chain. Moreover, because here we focus on the low-energy physics, only the overlap with the Majoranas at the junction are relevant. This extreme case allows us to simplify the problem and introduce a maximal junction asymmetry by choosing
\begin{equation}\label{sym_1}
	\phi_{ j,  \text{L}}^{} = \phi,
\end{equation}
and
\begin{equation}\label{sym_2}
	\phi_{ j,  \text{R}}^{} = 0,
\end{equation}
so that the phase an electron accumulates to go from the left to the right superconductor is~$ \phi_{J, \text{L}} - \phi_{1, \text{R}} = \phi$. In this limit, the Hamiltonian of the right chain~$\mybar{H}_{\text{K}, \text{R}} $ is identical to the Hamiltonian~${H}_{\text{K}, \text{R}}$ in Eq.~\eqref{Kitaev_Hamiltonian} since it is independent of the phase difference~$\phi$. In contrast, the Hamiltonian of the left chain takes the form
\begin{align}
		\mybar{H}_{\text{K}, \text{L}} (\phi)=  & -\mu \sum_{j}^{J} c_{j, \text{L}}^\dagger c_{j, \text{L}}^{} - t  \sum_{j}^{J - 1} \Big(  c^\dagger_{j+1, \text{L}} c_{j, \text{L}}^{} + \text{H.c.} \Big)  \nonumber \\	 &+ \Delta  \sum_{j}^{J - 1} \Big( e^{-2i \phi} \, c_{j+1, \text{L}}^{} c_{j,\text{L}}^{}  + \text{H.c.} \Big).
\end{align}
The tunneling Hamiltonian 
	\begin{equation}
		\mybar{H}_\text{T} = - \delta t \, \left(   c_{1, \text{R}}^\dagger c_{J,\text{L}}^{} +   \, c_{J, \text{L}}^\dagger c_{1, \text{R}}^{} \right)
	\end{equation}
is now independent of~$\phi$ since there is no phase drop across the weak link~$(\phi_{\delta t} = 0$). The unitary transformation that rotates the above Hamiltonian to the link basis is specified by the two operators
\begin{equation} \label{g_left}
	\mathcal{G}_\text{L}^{j, j^\prime}   = \delta_{j, j^\prime} \, \tau_z,
\end{equation}
and
\begin{equation} \label{g_right}
	\mathcal{G}_\text{R}^{j, j^\prime} = 0.
\end{equation}

Using the two operators~\eqref{g_left} and~\eqref{g_right}, we can derive an analytical expression for the coefficient~$\lambda$ in Eq.~\eqref{lambda_def},
\begin{equation} \label{lambda_upper}
	\lambda =  \mu  \rho^2 E_\text{M},
\end{equation}
with the density of states
\begin{equation}\label{key}
 \rho =	\dfrac{1}{\sqrt{4 t^2 - \mu^2} },
\end{equation}
and with~$E_\text{M}$ as the energy of the fractional Josephson effect, defined in Eq.~\eqref{EM}. For simplicity, let us first consider the limit of half-filling ($\mu=0$), a rather artificial limit for realistic semiconducting systems. Here, the coefficient~$\lambda$ is equal to zero, which can be understood as follows. At~$\mu = 0$, the Hamiltonian of an isolated Kitaev chain satisfies the particle-hole symmetry
\begin{equation} 
	\mathcal{H}^{j,j^\prime}_{\text{K}}  =	    (-1)^{j}  \tau_x \: 	\mathcal{H}^{j,j^\prime}_{\text{K}}   \: (-1)^{j^\prime}  \tau_x,
\end{equation}
which requires the spinors of each two degenerate states~$\ket{v}$ and~$\ket{u}$ to be related by
\begin{equation} \label{mu_symmetry}
	\ket{v}_n = (-1)^{j} \, \tau_x \, \ket{u}_n.
\end{equation}
Because of the symmetry~\eqref{mu_symmetry}, the contribution of above-the-gap states in Eq.~\eqref{lambda_def} sums to zero, even if the two chains are not identical, as long as each chain is long enough so that the zero modes decaying from its two ends do not overlap. This conclusion, in fact, holds regardless of the form of~$\mathcal{G}$ (\ie~for all profiles of the magnetic field).

More realistically, for~$n$-doped semiconducting nanowires, the chemical potential~$\mu$ is typically in the vicinity of~$-2t$~(\ie~$\mu \approx -2t + \delta \mu$). In this limit, the density of states is approximated by~$1/ \sqrt{4 t \delta \mu}$, and the coefficient~$\lambda$ can be cast into the form
\begin{equation}\label{key}
	\lambda \approx \dfrac{ E_\text{M} }{2 \delta \mu}.
\end{equation}
To compare the amplitude of the new term~$\hbar \dot{\phi} H_\lambda (\phi)$ with the Josephson term~$H_\text{M}(\phi)$, we define the parameter
\begin{equation}\label{key}
	\beta \equiv \dfrac{\hbar \dot{\phi} \lambda }{E_\text{M}} \approx  \dfrac{ \Delta}{\delta \mu}, 
\end{equation}
because the time-dependent drive~$\hbar \dot{\phi} $ is bounded by the superconducting gap~$2 \Delta$, otherwise the low-energy description would break down. Since typically~$\delta \mu > \Delta$, we expect that~$\beta \lesssim  1$\textemdash that is, the new term is comparable to but does not exceed the Josephson term~$H_\text{M}(\phi)$.

We note that the artificial case, defined by~Eqs.~\eqref{sym_1} and~\eqref{sym_2}, represents an upper bound on the parameter~$\lambda$ only for the field profiles treated in this section\textemdash that is, when the external magnetic field is localized within the superconducting loop. The coefficient~$\lambda$ is bounded because, for localized magnetic fields, the phase acquired at each site of the chain is guaranteed to be less than the total phase difference~(\ie~$|\phi_{j,\alpha}|\leq \phi$). When an electron takes a direct path\textemdash that is, a path that does not traverse a given site twice\textemdash from the left to the right superconductor, the acquired phase increases monotonically from zero at the left superconductor to~$\phi$ at the right one. This behavior does not necessarily hold for a magnetic field that is nonzero at the junction and inside the superconductors, which induces an additional current to expel the field from the bulk of the superconductor (the Meissner effect)~\cite{tinkham}. For such a nonlocalized field, we expect that more complicated phase profiles arise, which may be non-monotonic. That is, for a specified path between the two superconductors, the accumulated phase may overshoot above the total phase difference~$\phi$ and then decline to~$\phi$ at the end of the path, implying that the site phases~$|\phi_{j,\alpha}|$ are no longer bounded by~$\phi$. In Ref.~\cite{Riwar2021}, a related effect was predicted for regular Josephson-junction circuits: for a total phase drop~$\phi$ across an array of Josephson junctions, the phases at the individual junctions do not in general add monotonically (i.e., with the same sign) to the total value~$\phi$. In quantum circuit theory, such a non-monotonic behavior can be represented by an effective negative capacitance~\cite{Riwar2021}.

To summarize, we have investigated the scaling of the coefficient~$\lambda$ for a realistic device geometry and discussed its dependence on the parameters and the symmetry of the system. The novel correction term vanishes only for a system that is symmetric in all parameters, or for a band at perfect half-filling (\ie~zero chemical potential). Far away from half-filling and for an asymmetric system, the term is maximal. We have also provided an upper bound on~$\lambda$, assuming that the magnetic field is zero close to the superconductors. This bound can be exceeded for arbitrary profiles of the magnetic field.

\begin{figure}[t]
	\centering
	\includegraphics[scale=1]{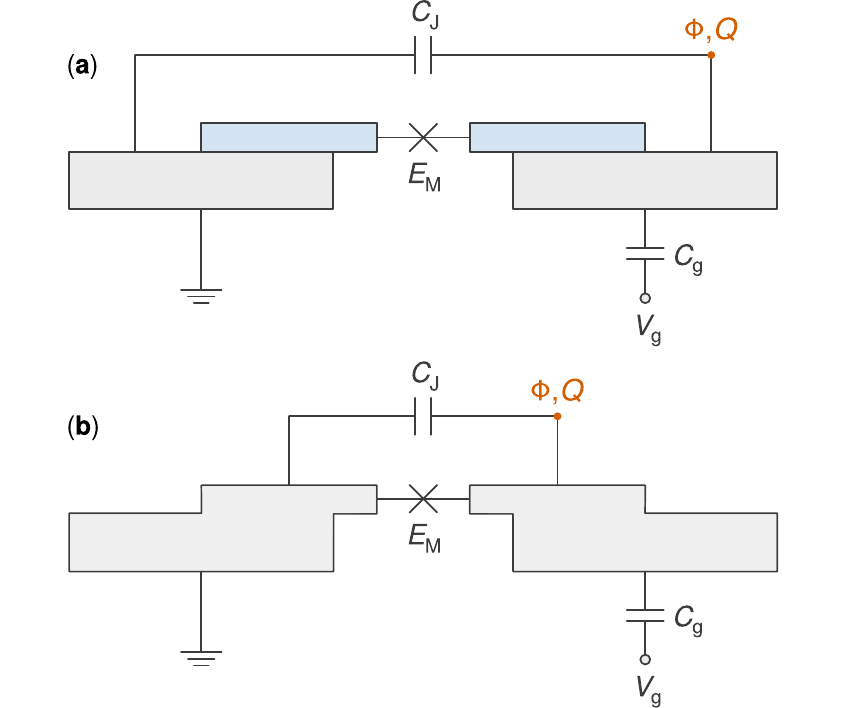}
	\caption{Circuit diagram consisting of a superconducting island connected to a grounded superconductor through a topological Josephson junction of capacitance~$C_\text{J}$ and energy~$E_\text{M}$. The island at node~$\Phi$ is connected to a gate voltage~$V_\text{g}$ via a capacitance~$C_\text{g}$. The two panels denote the two different basis\textemdash the irrotational and the link basis\textemdash for the loop geometry in Fig.~\ref{figure_model}.~(\textbf{a})~The chain (blue) and the superconductor (gray) are separate entities. In the loop geometry, the phase drop is distributed along the Kitaev chains as well as the weak link (\ie~the irrotational basis). In the open-circuit geometry, this choice corresponds to Hamiltonian~\eqref{circuit_ham} in which the charge on the island is shifted by the charge induced on the chains.~(\textbf{b})~The chains are considered a part of the metallic superconductor. In the loop geometry, the phase difference between the two superconductors is incorporated solely in the weak link (\ie~the link basis). In the open-circuit geometry, this choice corresponds to Hamiltonian~\eqref{circuit_ham_without}\textemdash that is,~$\lambda = 0$. The two Hamiltonians are related by the unitary transformation~\eqref{unitary_lamabda}. }
	\label{figure_circuit}
\end{figure}

\section{Circuit Hamiltonian for the fractional Josephson effect}\label{circuit}

In Sec.~\ref{current_measurement_sec}, we showed how the coefficient~$\lambda$ can be measured in a closed-loop geometry. Here, we examine the extent to which the nontrivial physics of surface charges extend to open circuits (\ie~cutting the superconducting loop to form a charge island controlled by a gate voltage, as depicted in Fig.~\ref{figure_circuit}). For the open circuit, the relevant dynamics stems from charging and discharging of a capacitor rather than a varying magnetic field.

Here, we revisit an assumption that we made in Sec.~\ref{irrotational_gauge_section} when calculating the irrotational vector potential. Namely, we assumed that the Kitaev chain itself is a weak conductor\textemdash that is, it fails to screen the electric field~$\vec{E}_{\dot{B}}$ in Eq.~\eqref{induced_E}, which implies that the induced charge~$eG$ and, in turn, the coefficient~$\lambda$ are nonzero. This assumption is realistic for Kitaev chains made of either semiconducting nanowires or topological insulators with low charge-carrier density. In the other extreme, the chains fully screen the electric field such that both~$eG$ and~$\lambda$ are zero. 

In this section, we compare these two extremes in the open-circuit geometry. First and in line with Sec.~\ref{irrotational_gauge_section}, we consider the chains irrelevant to the overall capacitance between the two superconductors~[Fig.~\ref{figure_circuit}(\textbf{a})], equivalent to no screening inside the chains~($\lambda \neq 0$). Second, we consider the chains a part of the bulk superconductors~[Fig.~\ref{figure_circuit}(\textbf{b})], equivalent to perfect screening inside the chains~($\lambda = 0$). Starting with the first extreme, the Lagrangian of the circuit in Fig.~\ref{figure_circuit}(\textbf{a}) can be written as~\footnote{The~$\lambda$ term in Eq.~\eqref{circuit_lagrang} is proportional to both the flux coordinate and its time derivative. It therefore can be treated as a potential or a kinetic energy term, which changes its sign in the Lagrangian. Here, we treat it as a potential energy term. Nonetheless, the sign convention does not alter the observables because they depend on~$\lambda^2$ [\eg~the charge fluctuations in Eq.~\eqref{fluctuation_lambda}]}
\begin{align} \label{circuit_lagrang}
	L_\text{c}^{\lambda} = &\,  \dfrac{ C_\text{J}}{2} \dot{\Phi}^2  + \dfrac{ C_\text{g}}{2} \big( \dot{\Phi} - V_\text{g} \big)^2 \nonumber \\ & - i \big( E_\text{M} + \lambda e \dot{\Phi}  \big) \, \cos \bigg( \dfrac{\varphi}{2} \bigg)  \: \gamma_3\gamma_2,
\end{align}
where the reduced flux~$\varphi \equiv 2 \pi \Phi / \Phi_0$ is the phase acquired by a Cooper pair such that~$\varphi = 2 \phi$ with~$\phi$ denoting the phase acquired by an electron as defined in Eq.~\eqref{tunneling_link_gauge}. The superconducting flux quantum~$\Phi_0$ equals~$h / (2e)$ and~$\Phi$ is the node flux of the island (Fig.~\ref{figure_circuit}). The gate voltage is coupled to the superconductor via a capacitance~$C_\text{g}$. Here, the coefficient~$\lambda$ is the same as that derived in the previous section because of the exact equivalence of the irrotational vector potential~\eqref{irrotational_vector_potential} and the electric field between two coplanar capacitor plates. This equivalence holds only because we assumed that the magnetic field is localized within the loop and does not touch the superconductors. For a generic magnetic field distribution, the coefficient~$\lambda$ is not the same for the loop and open-circuit geometries.

Following the standard quantization procedure~\cite{Devoret2017}, the resulting Hamiltonian is
\begin{align}\label{circuit_ham}
	H_\text{c}^\lambda = &~  4 E_\text{C} \bigg[   \op{n}   - n_\text{g}  - \dfrac{	i \lambda}{2}    \cos  \bigg( \dfrac{\ops{\varphi}}{2}  \bigg) \gamma_3^{} \gamma_2^{} \bigg]^2 \nonumber \\ & + i E_\text{M} \cos \bigg( \dfrac{\ops{\varphi}}{2} \bigg)  \gamma_3 \gamma_2.
\end{align}
The canonically conjugate variables obey the commutator
\begin{equation} 
	[\ops{\varphi}, \op{n}] = - i,
\end{equation}
where the number operator is defined as~$\op{n} \equiv \op{Q} / (-2e)$ with~$Q$ as the charge on the node in Fig.~\ref{figure_circuit}, conjugate to the flux coordinate~$\Phi$\textemdash that is,
\begin{equation} 
	Q \equiv \dfrac{\partial L_\text{c}^\lambda}{\partial \dot{\Phi}}.
\end{equation}
The charging energy~$E_\text{C} \equiv e^2 / (2 C)$ with
\begin{equation} 
	C \equiv C_\text{J} + C_\text{g}. 
\end{equation}
The gate voltage induces an offset charge~$n_\text{g} = C_\text{g} V_\text{g} / (2 e)$.

If, on the other hand, we assume that the chains can screen the electric field [Fig.~\ref{figure_circuit}(\textbf{b})], we obtain the standard Hamiltonian without the~$\lambda$ term
\begin{equation}\label{circuit_ham_without}
	{H}_\text{c} = 	 4 E_\text{C} \big(  \op{n} - n_\text{g} \big)^2 + i E_\text{M} \cos \bigg( \dfrac{\ops{\varphi}}{2} \bigg)\: \gamma_3 \gamma_2.
\end{equation}
The two Hamiltonians~\eqref{circuit_ham} and~\eqref{circuit_ham_without} are related by the unitary transformation
\begin{equation}\label{unitary_lamabda}
	U = e^{   {\lambda} \sin (\ops{\varphi} /2) \gamma_3 \gamma_2 },
\end{equation}
which transforms the number operator~$\op{n}$ as
\begin{equation} 
	U \op{n} U^\dagger = \op{n} - \dfrac{i \lambda}{2}   \cos \bigg( \dfrac{\ops{\varphi}}{2} \bigg)\: \gamma_3 \gamma_2 .
\end{equation}
The two Hamiltonians then have the same eigenvalues but different eigenfunctions, seeing that Hamiltonian~\eqref{circuit_ham} includes an additional charge proportional to~$\lambda$ induced on the Kitaev chains.

\begin{figure}[t]
	\centering
	\includegraphics[scale=1]{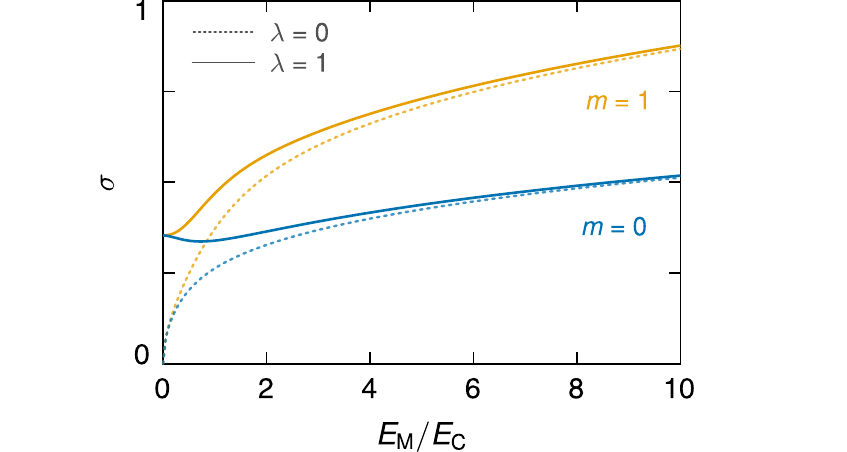}
	\caption{Fluctuations of the Cooper-pair number stored on the island as a function of the ratio~$E_\text{M} / E_\text{C}$ for the lowest two eigenvalues of~Hamiltonian~\eqref{circuit_ham}, labeled here by~$m$. The fluctuations are averaged over the charge offset~$n_\text{g}$. For small charging energies~$(E_\text{M} \gg E_\text{C}$), they scale with~$({E_\text{M} / E_\text{C}})^{1/4}$, and, in the basis of the Hamiltonian~$H_\text{c}^\lambda$, they are equal to~$\lambda / (2\sqrt{2})$ at zero Josephson energy~$(E_\text{M} = 0)$. Solid lines denote~$\lambda = 1$ and dotted lines~$\lambda = 0$.}
	\label{figure_fluctuation}
\end{figure}

The difference between the two Hamiltonians~$H_\text{c}$ and~$H_\text{c}^\lambda$ is most pronounced in the regime~$E_\text{C} \gg E_\text{M}$, where the standard fractional Josephson effect is suppressed, but the~$\lambda$ term survives. This limit corresponds to setting~$E_\text{M}$ to zero in Hamiltonians~\eqref{circuit_ham} and~\eqref{circuit_ham_without}. The two resulting Hamiltonians predict a radically different behavior for the fluctuations of the charge~$Q$ denoted in Fig.~\ref{figure_circuit}. While the Hamiltonian~$H_\text{c}$ commutes with the number operator~$\op{n}$ in the limit~$E_\text{C} \gg E_\text{M}$, the Hamiltonian~$H_\text{c}^\lambda$ does not. For this reason, charge fluctuations are not suppressed for the latter. Explicitly, using the eigenfunctions of the Hamiltonian~$H_\text{c} $, the variance of the operator~$\op{n}$ is
\begin{equation} \label{fluctuations_zero}
	\sigma^2 = \braket{\op{n}^2} - \braket{\op{n}}^2 = 0 .
\end{equation}
In contrast, using the eigenfunctions of~$H_\text{c}^\lambda$, it reads
\begin{equation}\label{fluctuation_lambda}
	\sigma_\lambda^2 = \Braket{\op{n}^2}_\lambda - \Braket{\op{n}}_\lambda^2 = \dfrac{\lambda^2}{8},
\end{equation}
where the subscript~$\lambda$ implies that the expected value is in the eigenbasis of Hamiltonian~\eqref{circuit_ham}. The eigenfunctions of the Hamiltonian~$H_\text{c}^\lambda$ are superpositions of adjacent charge states and, for an increasing~$\lambda$, they spread over an increasing number of charge states. For nonzero~$E_\text{M}$, we evaluate the variance numerically as a function of~$E_\text{M} / E_\text{C}$, as depicted in Fig.~\ref{figure_fluctuation}, by averaging it over the charge offset~$n_g$ to account for offset-charge drift~\cite{Serniak2019,Christensen2019}. The variance grows slowly as a function of the ratio~$E_\text{M} / E_\text{C}$ and scales with~$\sqrt{E_\text{M} / E_\text{C}}$ in the limit of small charging energies~(\ie~$E_\text{M} \gg E_\text{C}$). Even if~$E_\text{M}$ is vanishingly small, the number of Cooper pairs fluctuates by an amount commensurate with charge induced on the Kitaev chains.

The nature of these fluctuations can be understood based on the incompatibility of the transported charges and those stored in the capacitor. If the topological superconductor, namely the Kitaev chains, does not contribute to charge screening~($\lambda \neq 0$), then the capacitor plates consist of the conventional superconductor contacts which accept only integer number of Cooper pairs with charge~$2e$. In contrast, the fractional Josephson effect coherently transports charges in units of~$e$ because of the overlap between the Majorana bound states. Accordingly, in the limit of large charging energies, only integer Cooper-pair transport is energetically inhibited and charges in units of~$e$ may still tunnel across the weak link\textemdash that is, the fractional charges are figuratively stuck in limbo between the two superconducting contacts, thereby yielding nonzero fluctuations.

\section{Conclusions} \label{conclusions}

To conclude, we have explored time-dependent driving of Majorana-based quantum circuits and derived a low-energy Hamiltonian for the fractional Josephson effect in the regime of weak tunnel coupling. This Hamiltonian incorporates the electromotive force (emf) induced via a time-dependent magnetic field in the form of a correction term to the standard description of the fractional Josephson effect. This term depends on the geometry of the quantum circuit, the profile of the time-dependent magnetic field, and the (electrostatic) screening behavior of the topological superconductor. We have also provided simple measurement schemes for this new term, either via a current measurement in a loop geometry or via a charge-noise measurement in an open-circuit geometry.

Our results highlight the relevance of the unit of charge in quantum circuit theory. Here, the charge of the underlying condensate is~$2e$ and the charge transported across the weak link is~$e$. The behavior of the circuit relies on which of the two charges couple to the capacitor or to the time-varying magnetic field. This work opens the door for a more comprehensive study of the interplay between the emf and bound states (\eg~Andreev bound states in conventional $s$-wave junctions). The topologically trivial regime could unearth effects of even higher complexity because there are fewer symmetry-protected constraints and because the spectrum of the bound states cannot be detached from the quasiparticle continuum.

\section{Acknowledgments}
This work is supported by the Bavarian Ministry of Economic Affairs, Regional Development and Energy within Bavaria’s High-Tech Agenda Project ``Bausteine für das Quantencomputing auf Basis topologischer Materialien mit experimentellen und theoretischen Ansätze'' (grant No. 07 02/686 58/1/21 1/22 2/23). R.R. acknowledges funding from the German Federal Ministry of Education and Research within the program ``Photonic Research German'' (contract No.~13N14891). 

\appendix

\section{Schrieffer-Wolff transformation}\label{sw_appendix}
In this appendix, we derive the effective Hamiltonian~\eqref{effective_ham} of the fractional Josephson effect via a Schrieffer-Wolff transformation of the Hamiltonian~$ \mathcal{H}= \mathcal{H}_0 + \mathcal{H}_\text{T} + \hbar \dot{\phi} \mathcal{G}$, as defined in Sec.~\ref{low-energy-description}. Up to first order in the tunneling amplitude~$\delta t$ and the time-derivative of the phase~$\dot{\phi}$, the effective Hamiltonian in the single-particle picture reads
\begin{equation} \label{general_effective}
		\begin{split}
			\mathcal{H}_\text{eff} = \: & \mathcal{P} \mathcal{H}_0 \mathcal{P} + \mathcal{P} \mathcal{H}_\text{T}\mathcal{P} + \hbar \dot{\phi} \, \mathcal{P}  \mathcal{G} \mathcal{P}	\\ &- { \hbar \dot{\phi} \,  \mathcal{P} \bigg(    \mathcal{H}_\text{T} \dfrac{\mathcal{Q}}{\mathcal{H}_0}  \mathcal{G}   +  \mathcal{G}   \dfrac{\mathcal{Q}}{\mathcal{H}_0} \mathcal{H}_\text{T}   \bigg) \mathcal{P}},
		\end{split}
\end{equation}
where the operator~$ \mathcal{P} $ projects to the low-energy subspace, comprised of the four degenerate ground states of the left and right chains, and it can be expressed in the two-by-two subspace of the left and right chains as
\begin{equation} 
	\mathcal{P}  =   \begin{pmatrix}
		\mathcal{P}_\text{L} & 0 \\
		0 & \mathcal{P}_\text{R}
	\end{pmatrix} ,
\end{equation}
with
\begin{equation}\label{key}
	\mathcal{P}_\alpha =  \sum_{v = 0, \tilde{0}} \ket{v_\alpha} \bra{v_\alpha},
\end{equation}
where the subscript~$\alpha$ denotes either the left or the right chain. The projector~$\mathcal{Q}  = 1 - \mathcal{P} $. First, because the ground states are at zero energy, the diagonal term
\begin{equation} 
	 \mathcal{P}  \mathcal{H}_0 \mathcal{P}   = 0.
\end{equation}
Second, the tunneling term can be written as
 \begin{equation} \label{tunneling_first}
 	\mathcal{P}  \mathcal{H}_\text{T} \mathcal{P}  = \begin{pmatrix}
 	0 &  \mathcal{P}_\text{L} \mathcal{W} \, \mathcal{P}_\text{R}  \\
 		 \mathcal{P}_\text{R} \mathcal{W}^\dagger \, \mathcal{P}_\text{L}  & 0
 	\end{pmatrix} ,
 \end{equation}
with
\begin{align}\label{key}
 \mathcal{P}_\text{L} \mathcal{W} \, \mathcal{P}_\text{R}  = 	- E_\text{M} \cos \phi \, \big(  &\ket{0_\text{L}} \bra{0_\text{R}} + \ket{\to_\text{L}} \bra{0_\text{R}} \nonumber \\ &- \ket{0_\text{L}} \bra{\to_\text{R}} - \ket{\to_\text{L}} \bra{\to_\text{R}} \big).
\end{align}
In second quantization, the tunneling term takes the form
\begin{align}\label{kwwey}
		P H_\text{T} P &= \dfrac{1}{2} \psi^\dagger \,   	\mathcal{P}  \mathcal{H}_\text{T} \mathcal{P}  \,  \psi \nonumber  \\
		&= - E_\text{M}  \cos \phi  \,  \big(   d_{\text{M,L}}^\dagger  d_{\text{M,R}}^{} 	+   d_{\text{M,L}}^{}    d_{ \text{M,R}}^{}   + \text{H.c.} \big),
\end{align}
where the fermionic operators~$  d_{\text{M},\alpha}^\dagger$ and~$ d_{\text{M},\alpha}$ of the zero modes can be defined as
\begin{equation} 
	d_{\text{M},\alpha}^\dagger = \psi^\dagger_\alpha \ket{0_\alpha} = \bra{\tilde{0}_\alpha} \psi_\alpha,
\end{equation}
and
\begin{equation} 
	d_{\text{M},\alpha}^{}  = \psi^\dagger_\alpha \ket{\tilde{0}_\alpha} = \bra{{0}_\alpha} \psi_\alpha.
\end{equation}


In terms of the Majorana operators~$\gamma_{\{1-4\}}$ (Fig.~\ref{figure_model}), the fractional Josephson effect takes the standard form
\begin{equation} 
P H_\text{T} P = i E_\text{M}	 \cos \phi \: \gamma_{3}  \gamma_{2},
\end{equation}
where the Hermitian Majorana operators are related to the fermionic operators by
\begin{equation} \label{majorana_1_op}
	d_\text{M,L}^{} = \dfrac{ \gamma_2 + i \gamma_1}{2},
\end{equation} 
and
\begin{equation} \label{majorana_2_op}
	d_\text{M,R}^{}  = \dfrac{ \gamma_4 + i \gamma_3}{2}.
\end{equation} 
In terms of the parameters of the Kitaev chain, the energy
\begin{equation} \label{EM}
		E_\text{M} = \dfrac{\delta t}{2} \,  \Big( {1 - Y^2  } \Big)  \bigg(1 - \dfrac{\mu^2}{4 t^2}\bigg),
\end{equation}
with
\begin{equation} 
	Y \equiv  \dfrac{ t - \Delta     }{ t + \Delta} .
\end{equation}

Third, we focus on the term~$ \mathcal{P} \mathcal{G} \mathcal{P} $. As discussed in Sec.~\ref{irrotational_gauge_section}, the matrix~$ \mathcal{G}_\alpha$ has the two-by-two structure
\begin{equation} \label{G_sub}
	\mathcal{G}_\alpha^{j, j^\prime} = \delta_{j,j^\prime} \: \dfrac{\phi_{j, \alpha}}{\phi}   \begin{pmatrix}
	1 & 0 \\
		0 & - 1 
	\end{pmatrix},
\end{equation}
The profile of the time-dependent magnetic field\textemdash that is, the distribution of the phase drop along the chains\textemdash is captured by the site phases~$\phi_{j, \alpha}$ defined in Eq.~\eqref{general_phase_symmetric}. We can write the term~$ \mathcal{P} \mathcal{G} \mathcal{P} $ as
 \begin{equation} 
	\mathcal{P} \mathcal{G} \mathcal{P}  = \begin{pmatrix}
		  \mathcal{P}_\text{L} \mathcal{G}_\text{L} \mathcal{P}_\text{L} &0  \\
	0 &  \mathcal{P}_\text{R} \mathcal{G}_\text{R} \mathcal{P}_\text{R} 
	\end{pmatrix} .
\end{equation}
The contribution of chain~$\alpha$ is
\begin{align}\label{key}
	 \mathcal{P}_\alpha \mathcal{G}_\alpha \mathcal{P}_\alpha = 2 g_\alpha^{}  \, \big( \ket{0_\alpha} \bra{0_\alpha} - \ket{\to_\alpha} \bra{\to_\alpha} \big),
\end{align}
with
\begin{equation}\label{key}
	g_\alpha^{} \equiv \dfrac{\bra{0_\alpha} \mathcal{G}_\alpha \ket{0_\alpha }}{2} = - \dfrac{\bra{\to_\alpha} \mathcal{G}_\alpha \ket{\to_\alpha }}{2}.
\end{equation}
In second quantization, the term~$ PGP $ takes the form
\begin{align}\label{key}
	PGP &= \dfrac{1}{2} \psi^\dagger \mathcal{P} \mathcal{G} \mathcal{P} \psi \nonumber \\
	&= \sum_\alpha  g_\alpha \, \big( d_{\text{M},\alpha}^\dagger  d_{\text{M},\alpha} - d_{\text{M},\alpha} d_{\text{M},\alpha}^\dagger  \big) .
\end{align}
Substituting with the Majorana operators in Eqs.~\eqref{majorana_1_op} and~\eqref{majorana_2_op} leads to the final form
\begin{equation}\label{key}
PGP =	 i g_\text{L}^{} \, \gamma_2 \gamma_1 + i g_\text{R}^{} \, \gamma_4 \gamma_3,
\end{equation}
where the coefficients~$ g_\text{L} $ and~$ g_\text{R} $ depend on the profile of the magnetic field in the left and right chains, respectively. Importantly, the term~$PGP$ is approximately zero because the coefficient~$ 	g_\alpha $ is exponentially suppressed, which can be understood as follows. The two ground states of the Kitaev chain can be expressed as a superposition of the two modes decaying from the left~$(\ell)$ and right~$(r)$ ends of the same chain as
\begin{equation}\label{zero_mode}
	\ket{0_\alpha} = \dfrac{\ket{0_\alpha^r} + \ket{0_\alpha^\ell}}{\sqrt{2}},
\end{equation}
and
\begin{equation}\label{zero_mode_tilda}
		\ket{\to_\alpha} = \dfrac{\ket{0_\alpha^r} - \ket{0_\alpha^\ell}}{\sqrt{2}}.
\end{equation}
Based on the two-by-two structure~\eqref{G_sub} of~$\mathcal{G}_\alpha$, the matrix elements~$	\bra{0^\ell_\alpha}   \mathcal{G}_\alpha   \ket{0^\ell_\alpha}$ and~$	\bra{0^r_\alpha}   \mathcal{G}_\alpha   \ket{0^r_\alpha} $ equal zero. Moreover, for a chain that is long enough so that the zero modes decaying from its two ends do not overlap, the matrix element~$	\bra{0^\ell_\alpha}   \mathcal{G}_\alpha   \ket{0^r_\alpha}$ is exponentially suppressed. For example, for a chain of odd number of sites~$J$ and with~$\mu =0$, the coefficient~$g_\alpha$ takes the analytical form
\begin{equation}\label{key}
g_\alpha = 	\dfrac{1}{2} \,  \big( {1 - Y^2  } \big) \, Y^{(J-1)/2} \,  \sum_{j \in \text{odd}}  \dfrac{\phi_{j,\alpha}}{\phi},
\end{equation}
which is exponentially suppressed as~$J$ increases.

Finally, analogous to the term~$PH_\text{T} P$, the last term
\begin{equation}\label{key}
	\Gamma \equiv PH_\text{T}\dfrac{Q}{H_0} G P + P G \dfrac{Q}{H_0}H_\text{T} P
\end{equation}
can be cast into the form
\begin{equation}
	\Gamma = i \, \big(    h_{42} \, \gamma_4 \gamma_2 +   h_{31} \, \gamma_3 \gamma_1 +  h_{32} \, \gamma_3 \gamma_2   +  h_{41} \, \gamma_4 \gamma_1 \big),
\end{equation}
which highlights the coupling between the four Majoranas of the left and right chains, as depicted in Fig.~\ref{figure_model}(\textbf{b}). The coupling coefficients can be written as
\begin{align}
	 h_{42}  &\equiv -  \dfrac{\Im \{ \beta_{00} + \beta_{0 \to}  \}}{2} ,\\
	 h_{31} &\equiv -  \dfrac{\Im \{ \beta_{00} - \beta_{0 \to}  \}}{2} ,\\
	 h_{32} &\equiv -  \dfrac{\Re \{ \beta_{00} - \beta_{0 \to}  \}}{2}, \\
	 	 h_{41} &\equiv  \: \dfrac{\Re \{ \beta_{00} + \beta_{0 \to}  \}}{2} , 
\end{align}
with the matrix element
\begin{align}\label{key}
	\beta_{pq} \equiv & \sum_{v_\text{R} \neq 0,\to} \,  \epsilon_{\vr}^{-1} \,   \bra{p_\text{L}} \mathcal{W} \ket{v_\text{R}} \bra{v_\text{R}} \mathcal{G}_\text{R} \ket{q_\text{R}} \nonumber \\
	&+  \sum_{v_\text{L} \neq 0,\to} \epsilon_{\vl}^{-1}  \,  \bra{p_\text{L}} \mathcal{G}_\text{L} \ket{v_\text{L}} \bra{v_\text{L}} \mathcal{W} \ket{q_\text{R}} .
\end{align}
The form~\eqref{w_structure} of~$\mathcal{W}$ implies that the coefficients~$h_{42}$ and~$h_{31}$ are proportional to~$\sin \phi$ (because they are proportional to the imaginary part of~$\beta_{pq}$), while~$h_{32}$ and~$h_{41}$ are proportional to~$ \cos \phi$. In this work, we only consider long Kitaev chains where the modes at the two ends of each chain do not overlap. Consequently, the coefficients~$h_{42}$,~$h_{31}$, and~$h_{41}$ are exponentially suppressed, and only~$h_{32}$ survives. To illustrate this behavior, we represent the zero-energy ground states as a superposition of the left and right modes, as in Eqs.~\eqref{zero_mode} and~\eqref{zero_mode_tilda}. For example, the coefficient~$h_{42}$ can be simplified to
	\begin{equation} 
		h_{42} =  - \delta t \sin \phi \, \sum_{\vr  > 0} \, \epsilon_{\vr}^{-1} \, {\bra{0_\text{L}^r} \mathcal{I} \ket{\vr} \bra{\vr} \mathcal{G}_\text{R} \ket{0_\text{R}^r}},
	\end{equation}
where
\begin{equation}\label{key}
	\mathcal{I}_{j,j^\prime} =  \delta_{j,J} \:  \delta_{j^\prime,{1}}  \: \begin{pmatrix}
		1 & 0 \\0 & 	1
	\end{pmatrix}.
\end{equation}
The coefficient~$h_{42} $ couples the right end of the left chain to the right end of the right chain and depends on the interaction of the magnetic field with the right chain via the operator~$\mathcal{G}_\text{R}$. The coefficient~$h_{32}$ reduces to
\begin{align} 
	h_{32} =  ~ &  \delta t   \cos \phi \, \Bigg[ \sum_{\vr >0 }  \epsilon_{\vr}^{-1} \, \bra{0_\text{L}^r}  \mathcal{Z} \ket{\vr} \bra{\vr} \mathcal{G}_\text{R} \ket{0_\text{R}^\ell} \nonumber \\ & +  \sum_{\vl > 0}  \epsilon_{\vl}^{-1}  \, \bra{0_\text{R}^\ell}  \mathcal{Z}^\dagger  \ket{\vl} \bra{\vl} \mathcal{G}_\text{L} \ket{0_\text{L}^r} \Bigg]  ,
\end{align}
with
\begin{equation}\label{key}
	\mathcal{Z}_{j,j^\prime} =  \delta_{j,J} \:  \delta_{j^\prime,{1}}  \: \begin{pmatrix}
	1 & 0 \\0 & 	-1
	\end{pmatrix},
\end{equation}
and it couples the two adjacent ends of the left and right chains. 

Based on Eq. \eqref{general_effective}, the effective Hamiltonian takes its final form
\begin{equation} 
H_\text{eff} = 	i E_\text{M}	 \cos \phi \: \gamma_{3}  \gamma_{2} - i \hbar \dot{\phi} \, {h_{32}} \: \gamma_{3}  \gamma_{2}.
\end{equation}

\section{Linear response of the current measured across the weak link}\label{sus_appendix}

This appendix derives an expression for the susceptibility in response to fluctuations in the phase drop across the weak link that couples two Kitaev chains. Specifically, we calculate the linear response of the current~$\mybar{I}_0$ across the weak link to fluctuations in the form~$ \phi (t) \rightarrow  \phi + \delta \phi (t) $. Up to first order in the fluctuations, the Hamiltonian in the irrotational basis can be approximated by
\begin{equation} 
	\mybar{H}(\phi + \delta \phi) \approx \mybar{H}(\phi) - \dfrac{\hbar}{e}  \delta \phi(t) \mybar{I}.
\end{equation}
The time-dependent Kubo formula in the frequency domain can be written as
\begin{equation} 
	\braket{\mybar{I}_0(\omega)} =   	\braket{\mybar{I}_0(\omega)}_0 + \delta \phi(\omega)\,  \chi (\omega),
\end{equation}
with the Fourier transform of the fluctuations as
\begin{equation} 
	\delta \phi (\omega)  =  \int_{-\infty}^{\infty} \mathrm{d}t \:  e^{i \omega t} \delta \phi(t).
\end{equation}
The susceptibility is defined as
\begin{equation} \label{sus_in_app}
		\chi^{}(\omega) =  \dfrac{i}{e} \int_{-\infty}^{\infty} \mathrm{d}t \: e^{i\omega t } \, \Theta(t) \, \Braket{[ \mybar{I}_0(0), \mybar{I}(-t)  ]  }_0,
\end{equation}
where~$\Theta(t)$ denotes the Heaviside step function. Using the current~$\mybar{I}$ in the form~\eqref{current_irrotational}, the susceptibility can be decomposed into
\begin{equation} 
	\chi^{}(\omega)  = 	\chi^{}_0(\omega)  + 	\chi^{}_G(\omega) ,
\end{equation}
where
\begin{equation} 
	\chi^{}_0(\omega) =  \dfrac{i}{e} \int_{-\infty}^{\infty} \mathrm{d}t\: e^{i\omega t } \, \Theta(t) \, \Braket{[ \mybar{I}_0(0), \mybar{I}_0(-t)  ]  }_0,
\end{equation}
and
\begin{equation} 
	\chi^{}_G(\omega)	= i \int_{-\infty}^{\infty} \mathrm{d}t\: e^{i\omega t } \, \Theta(t) \, \Braket{[ \mybar{I}_0(0), \dot{\mybar{G}}(-t)  ]  }_0.
\end{equation}
In line with the low-energy approximation in Sec.~\ref{low-energy-description}, only terms that are first order in the tunneling amplitude are kept. We therefore drop the contribution~$	\chi^{}_0(\omega)$ and approximate the time evolution of~$\mybar{G}$ by
\begin{equation} 
	\dot{\mybar{G}}(-t) \approx  \dfrac{i}{\hbar}  \Big[H_0, e^{-i \mybar{H}_0 t / \hbar} \: \mybar{G} \: e^{ i \mybar{H}_0 t / \hbar}     \Big] .
\end{equation}
Instead of evaluating the matrix element in the above expression for states belonging to the relevant low-energy subspace, one can define a susceptibility operator that acts on the low-energy subspace by replacing the expectation value~$\braket{\bullet}$ in Eq.~\eqref{sus_in_app} by~$ \mybar{P} \bullet \mybar{P}$, where the operator~$\mybar{P}$ projects onto the desired subspace.

Substituting with~$\dot{\mybar{G}}$ into the operator form~\eqref{sus_op} and integrating leads to
\begin{equation} 
	\op{\overline{\chi}} (\omega) = - \mybar{P} \bigg(   \mybar{I}_0 \mybar{Q} \dfrac{i \mybar{H}_0}{{ \hbar \omega }  - \mybar{H}_0} \mybar{G} + \mybar{G} \dfrac{i \mybar{H}_0}{{ \hbar \omega }  - \mybar{H}_0}  \mybar{Q}  \mybar{I}_0  \bigg) \mybar{P}.
\end{equation}
The susceptibility operator~$	\op{\overline{\chi}} (\omega) $ is non-Hermitian and hence its expected value has real and imaginary parts. It can therefore be decomposed into two operators that correspond to the real and imaginary parts of the susceptibility, as in Eqs.~\eqref{hermitian_part} and~\eqref{antihermitian_part}. In the low-frequency limit, the operator corresponding to the real part of the susceptibility can be simplified to
\begin{equation} \label{hermitian_part_app}
\op{\overline{\chi}}_r (\omega)  = i \mybar{P} \left(   \mybar{I}_0 \mybar{Q}  \mybar{G} - \mybar{G}  \mybar{Q}  \mybar{I}_0  \right) \mybar{P},
\end{equation}
and that corresponding to the imaginary part to
\begin{equation} 
	\op{\overline{\chi}}_i (\omega) =  { \hbar  } \omega \mybar{P} \bigg(   \mybar{I}_0  \dfrac{\mybar{Q}}{ \mybar{H}_0} \mybar{G} + \mybar{G} \dfrac{ \mybar{Q} }{ \mybar{H}_0}   \mybar{I}_0  \bigg) \mybar{P}.
\end{equation}
As discussed in Sec.~\ref{low-energy-description}, the imaginary part of the susceptibility is related to the current that arises from the interaction of the time-dependent external magnetic field with the Kitaev chain. In this Appendix, we focus on the real part of the susceptibility. In the link basis, the operator~\eqref{hermitian_part_app} can be cast into the differential form
\begin{equation} \label{real_sus_differential}
		\op{{\chi}}_r^{} =  \dfrac{e}{ i\hbar} \, \partial_\phi \big(  P  H_\text{T} {Q}  {G} P - P {G}  {Q}  H_\text{T} P \big),
\end{equation}
because only the current~$I_0$ depends on~$\phi$. Analogous to Appendix~\ref{sw_appendix}, the bracketed term simplifies to
\begin{equation} 
	P  H_\text{T} {Q}  {G} P - P {G}  {Q}  H_\text{T} P = \, \dfrac{h^\prime_{32}}{2} \gamma_{3} \gamma_{2},
\end{equation}
where rest of the coefficients are exponentially suppressed. The coefficient~$h^\prime_{32}$ reads
\begin{equation} 
	h^\prime_{32} = \Im \big\{  \bra{0_\text{L}} \mathcal{W}   \Xi_\text{R}  \mathcal{G}_\text{R} \ket{0_\text{R}}  -  \bra{0_\text{L}}   \mathcal{G}_\text{L} \Xi_\text{L}  \mathcal{W} \ket{0_\text{R}}  \big\} ,
\end{equation} 
with
\begin{equation} 
	\Xi_\alpha \equiv  \sum_{v_\alpha > 0} \big( \proj{v_\alpha} + \proj{\tilde{v}_\alpha} \big).
\end{equation}
Using definition~\eqref{w_structure} for the tunneling matrix~$\mathcal{W} $ and definition~\eqref{G_sub} for~$\mathcal{G}_\alpha$, the coefficient~$h^\prime_{32}$ reduces to
\begin{equation} 
	h^\prime_{32} =  \dfrac{\phi_{1, \text{R}}^{} - \phi_{J, \text{L}}^{} }{\phi}   \, E_\text{M} \, \sin \phi ,
\end{equation} 
where the prefactor is independent of the phase~$\phi$ since the phases~$  \phi_{1, \text{R}}^{}$ and~$ \phi_{J, \text{L}}^{} $ are proportional to~$ \phi $ [\cf~Eq.~\eqref{general_phase_symmetric}]. 

Substituting into Eq.~\eqref{real_sus_differential} gives the final form
\begin{equation} 
\op{{\chi}}_{r}^{}  = \dfrac{e}{\hbar} \bigg( 1 - \dfrac{\phi_{\delta t}}{\phi} \bigg)  \, i E_\text{M}  \cos \phi \, \gamma_{3} \gamma_{2},
\end{equation}
where~$\phi_{\delta t}$ is the phase drop across the weak link in the irrotational basis~$\mybar{H}$, and~$\phi$ is the phase drop across the weak link in the link basis~$H$ (\ie~the overall phase difference between the two superconductors).

\section{Current in the irrotational gauge}\label{current_irrotational_appendix}
In this appendix, we derive the current in Eq.~\eqref{current_irrotational}. In the basis of the Hamiltonian~$\mybar{H}$, one can define the current as
\begin{equation}
	\mybar{I} = - \dfrac{e}{\hbar} \, \partial_\phi \mybar{H}(\phi).
\end{equation}
Rotating the Hamiltonian to the link basis~($\mybar{H} = U H U^\dagger $) and differentiating with respect to~$\phi$ results in
\begin{align}\label{key}
		\mybar{I} = - \dfrac{e}{\hbar}  \, \Big[ (\partial_\phi U) H U^\dagger +   U (\partial_\phi H )U^\dagger  +   U H (\partial_\phi U^\dagger)  \Big].
\end{align}
Rearranging the terms and using the form~\eqref{current_link} of the current across the weak link yields
\begin{equation}\label{key}
			\mybar{I}  = U I_0 U^\dagger  - \dfrac{e}{\hbar} \, \Big[ (\partial_\phi U) H U^\dagger +    U H (\partial_\phi U^\dagger)  \Big].
\end{equation}
Rotating back to the basis of the Hamiltonian~$ \mybar{H} $ leads to
\begin{equation}\label{key}
		\mybar{I}   = \mybar{I}_0 - \dfrac{e}{i\hbar} \, \Big[ i (\partial_\phi U)  U^\dagger \mybar{H} +    \mybar{H} iU (\partial_\phi U^\dagger)  \Big].
\end{equation}
In the irrotational basis, the operator~$ G $ in Eq.~\eqref{G_definition} can be written as
\begin{equation}\label{key}
	\mybar{G} = - i (\partial_\phi U)  U^\dagger = i U (\partial_\phi U^\dagger ).
\end{equation}
Therefore, the current in the irrotational basis takes the final form
\begin{align}\label{key}
	\mybar{I} &= \mybar{I}_0 + \dfrac{e}{i\hbar} \big(  \mybar{G} \mybar{H} - \mybar{H} \mybar{G} \big) \nonumber \\
	&= \mybar{I}_0 + e \dot{\mybar{G}},
\end{align}
with
\begin{equation}\label{key}
	\dot{\mybar{G}} = \dfrac{i}{\hbar }  \big[ \mybar{H}, \mybar{G} \big].
\end{equation}

\newpage
\bibliography{refLib}

\end{document}